\documentclass[a4paper]{spie}  
\usepackage[]{graphicx}

\title{KMOS Data Flow: Reconstructing Data Cubes in One Step}

\author{Richard Davies, 
Alex Agudo Berbel, 
Erich Wiezorrek, 
Thomas Ott, 
and\\
Natascha M. F\"orster Schreiber
\skiplinehalf
Max Planck Institut f\"ur extraterrestrische Physik, 
 Postfach 1312, Garching 85741, Germany
}

\authorinfo{Further author information: (Send correspondence to
  R.D.)\\E-mail: davies@mpe.mpg.de}

\newcommand{\arcsec}{\hbox{$^{\prime\prime}$}}
\newcommand{\arcmin}{\hbox{$^{\prime}$}}
\newcommand{\micron}{\,\hbox{$\mu$m}}

\begin{document} 
  \maketitle 

\begin{abstract}

KMOS is a multi-object near-infrared integral field spectrometer with
24 deployable pick-off arms. Data processing is inevitably complex. We
discuss specific issues and requirements that must be addressed in the data reduction pipeline, the calibration, the raw and processed data formats, and the simulated data. 
We discuss the pipeline architecture. We focus on its modular style and show how these modules 
can be used to build a classical pipeline, as well as a more advanced
pipeline that can account for both spectral and spatial flexure as well
as variations in the OH background. A novel aspect of the pipeline is
that the raw data can be reconstructed into a cube in a single step. We
discuss the advantages of this and outline the way in which we have
implemented it.
We finish by describing how the QFitsView tool can now be used to
visualise KMOS data.
\end{abstract}



\section{KMOS Overview}
\label{sec:overview}

KMOS\cite{sha10a,sha10b} is a fully cryogenic seeing limited multi-object integral field spectrometer,
which is being built by a consortium of German and British institutes
as one of the second generation VLT instruments.
It is equipped with 24 integral field units, each with a
2.8\arcsec$\times$2.8\arcsec\ field of view, that can be deployed
on-the-fly using robotic arms anywhere within a 7\arcmin\ patrol
field.
The number and size of the IFUs, and the spectral resolution, have
been chosen according to the key science driver, to study the morphology and kinematics of emission lines in high redshift galaxies.
Each of the 3 identical segments in the instrument comprises 8
pick-off arms (including K-mirrors to keep the fields oriented
correctly on sky), filter wheels, the integral field unit (that
uses mirrors to slice the image and re-arrange it along a
pseudo-slit), and a spectrograph with a 2k$\times$2k HAWAII-2RG detector.
A number of gratings are available, allowing observations across
various bands from 0.8--2.5\micron\ at a resolution of $R\sim3500$ in
order to enable science observations to probe between the bright
atmospheric OH lines.
An additional grating enables observation of the H and K bands
simultaneously at a lower resolution.
Since the pixel scale is nominally 0.2\arcsec, a single exposure with
KMOS generates 336 slit spectra, each about 2000 pixels in the
spectral direction and 14 pixels wide spatially. 
Each slit spectrum is separated from its neighbours with a gap of a
few blank pixels.
Although there are a large number of spectra, the absolute data volume
is not excessive: in a typical night the instrument will produce only
$\sim$5\,Gbyte of raw data (although it could be up to $\sim$50\,Gbyte in
the extreme case of short exposures).
The pipeline modules (recipes) are written in C using the ESO
Common Pipeline Library.
They can be executed using the standard ESO pipeline tools, either in the
command line with {\em EsoRex} or with the {\em Gasgano} GUI.

KMOS is currently in its manufacturing phase, and has already had its
technical first light\cite{sha10a,sha10b}.
The preliminary acceptance in Europe and first light on sky at the
VLT in Chile are both expected during 2011.
Further details on specific aspects of the instrument, including
hardware, software, control, and testing, can be found elsewhere in
these proceedings.

\section{Instrument Configuration and Observing Modes}

KMOS itself is opto-mechanically a very complex instrument that slices
images from 24 deployable pick-off arms in order to perform integral
field spectroscopy (also known as hyperspectral imaging).
However, from an observer's perspective, configuring and using the
instrument is rather straightforward.
Unlike many other integral field spectrometers, there is no choice of
pixel scale.
Neither is it necessary to adjust the centre of the spectral bandpass, or
optimise the slit width against resolution as can be the case for
longslit spectrometers.
And in contrast to most `workhorse' instruments, there are no
additional options such as imaging or polarimetry.
The only complex operation, allocating the 24 arms to specific science
targets in the patrol field, is greatly assisted by an automated tool
KARMA\cite{weg08} that optimizes the assignment taking into account
optical and mechanical constraints, as well as target priorities.
This simplicity in terms of configuration has a direct impact on the
data processing, since each set of frames can be processed in
essentially the same way.
However, there are more subtle features, related to the observing
modes\cite{weg08}, that need to be considered.

One of the observing modes that is expected to be used frequently is
`nod to sky'.
Here, 24 targets are observed in every pair of pointings, with most of
the arms on targets at the first position, and the remainder on
targets during the second exposure after a nod.
In this case, the sky background is removed simply by subtracting
alternate exposures for each arm.
But this means it is not always possible to classify an entire
exposure as `sky' or `object' since some arms may be on sky and others
on targets.
Consequently, classifications are made separately for each arm and
the data for each IFU are processed independently.

With the `stare' mode, the arms are on the same science target for
multiple sequential exposures.
For this mode, the observer has the option of occasionally observing a
sky pointing -- in which case that can be used to remove the background.
Alternatively, if the target is compact enough that it
can be dithered within the IFU's field of view, the background can be
recovered from sky pixels around the science target.
Another option, potentially one of the most efficient and effective
uses of the `stare' mode, would be to have a single dedicated sky arm, so
that the other 23 could always be on targets.
Ideally, one would subtract this sky exposure from all the other arms
-- with the advantage that it has monitored the background
(particularly the highly variable atmospheric OH lines) {\em during} the
science  exposure and so provides the best background measurement.
However, there are two issues that mean this method can only be applied
in a rather restricted sense:
(i) the effect of spatial and spectral curvature means that one must match
corresponding arms in each segment (e.g. arms 1, 9, and 17), since only
then do the detailed spectral profiles of the OH lines match exactly;
(ii) the IFUs have differing orientations on sky so that, for example,
the data for IFU\,9 would need to be rotated by 180$^\circ$, and that
for IFU\,17 flipped, before they can subtracted from IFU\,1.

The third mode is `mapping' in which the arms are positioned as closely
as possible and, with a sequence of 9 or 16 pointings, are used to map a
large (arcmin scale) contiguous field.
This imposes a requirement that the data processing software be able to
handle a large number of cubes: 24 IFUs and several exposures at each
of 16 pointings leads to in excess of 1000 data cubes that must
eventually be combined.

\section{Calibration}

\begin{figure}
\begin{center}
\includegraphics[height=5.5cm]{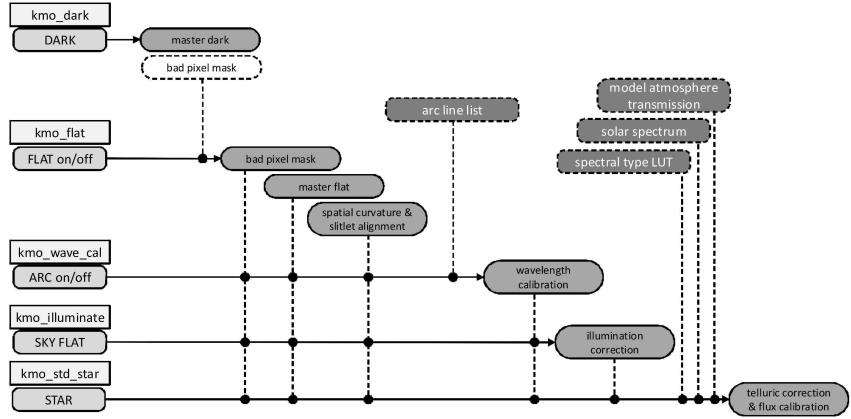}
\end{center}
\caption{ \label{fig:cascade} 
Calibration Cascade showing how the various calibration recipes are
associated, what input data they require (both from the instrument and
externally), and what calibrations they produce.
}
\end{figure}

Various calibrations are required to process KMOS data.
The way in which they are associated, the input data
required, and the output produced, are summarised in
Fig.~\ref{fig:cascade}.
The specific recipes or modules in this calibration cascade are
described below.
In order to make sure that the various calculations and fits performed
are robust, we employ a routine that flags and rejects deviant
values very efficiently, even when there are  only 4--5 values in a
sample.
The key aspect is to begin with an estimate of the standard
deviation based on a percentile clipping, and then improve on this
with a few iterations.

\subsection{Calibration Recipes}

\begin{figure}
\begin{center}
\includegraphics[width=15cm]{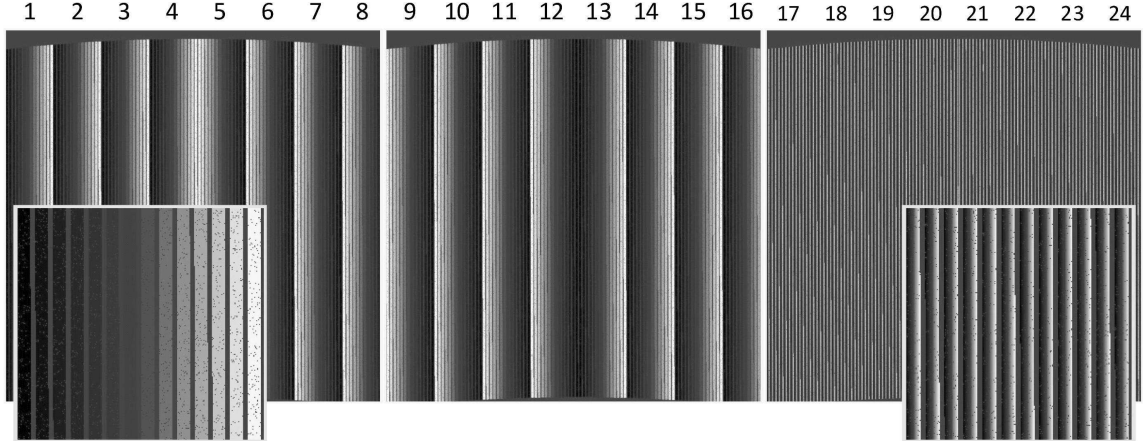}
\end{center}
\caption{ \label{fig:xcal} 
The calibration frames for XCAL in which the value of each
pixel indicates the spatial distance along the X-axis (right
ascension) of the IFU. The numbers along the top indicate to which IFU
each section of data belongs. 
Each IFU is arranged in 14 separate 2D slit spectra.
For the first two detectors the IFUs are oriented vertically on sky;
so each slitlet has a single value and values increase/decrease
between slitlets (left inset).
The 3rd detector appears different because the orientation of the IFUs
is 90$^\circ$ offset and so values increase across each slitlet
(right inset).
}
\end{figure}

\begin{description}

\item{\em kmo\_dark} combines several raw dark frames, and in the
process makes a first pass at identifying bad pixels. If the
exposure time is very short, it also yields the read noise. And for
very long exposures it provides a measure of the dark current.\vspace{-1mm}

\item{\em kmo\_flat} has two functions.
The first is to combine several `lamp on' and `lamp off' frames to make a
flatfield, in the process identifying additional bad pixels.
For KMOS, a bad pixel is defined as one that should be
ignored (for any reason) during the cube reconstruction.
As such, it also includes pixels that are simply not illuminated.
The second function is to trace the edges of each
slitlet in each IFU, and hence map the spatial curvature.
This is a key process, and is implemented in a robust way with no 
prior assumptions about the location of the slitlet edges.
As such, it works even if several IFUs are out of action.
The routine creates a pair of calibration frames called XCAL (see
Fig.~\ref{fig:xcal}) and YCAL in which the value of pixel denotes its
spatial location with respect to the centre of its IFU along the
right ascension and declination axes respectively.
Each pixel is also tagged with the identification of the IFU to which it
belongs.\vspace{-1mm}

\item{\em kmo\_wave\_cal} calculates the wavelength of each pixel,
treating each slitlet completely independently.
The initial identification of arc lines within a slitlet is done by
comparing every possible pair of line separations with those from a
linelist. With only the approximate scaling as a prior assumption,
this is a robust method that yields an unambiguous match between the
data and line list.
From this point, the lines are traced across each slitlet and hence
the wavelength at every pixel can be interpolated.
These are written into a frame called LCAL.\vspace{-1mm}

\item{\em kmo\_illuminate} is a spatial sky flat. 
It measures variations in the illumination across the field of
each IFU in order to correct for vignetting, either at the edge of the
patrol field or if an arm is in the shadow of another.\vspace{-1mm}

\item{\em kmo\_std\_star} is a standard recipe that both derives the
atmospheric and instrument transmission as a function of wavelength, and also
calculates the scaling factor for flux calibration.
The crucial issue here is to remove the spectral features.
This is assisted by temporarily dividing out a model atmosphere.
One then divides out a solar spectrum that has been
convolved to the same resolution (or, for AB stars observed in the
K-band, interpolate across Br$\gamma$).
Afterwards, the model atmosphere is multiplied back in.
A look-up table enables the code to estimate the stellar temperature
and remove the intrinsic spectral slope.

\end{description}

\subsection{Quality Control}

Monitoring the consistency and trends in the quality control (QC)
parameters provides  
information about the status and health of the instrument.
These parameters show that the instrument is functioning
correctly and that data can be calibrated properly (but do not
necessarily reflect the quality of the data itself).
Rather than list them all explicitly, a summary of the main
classes, the recipe that generates them, and their purpose is given:

\begin{tabular}{p{2.2cm}p{4.4cm}p{8.9cm}}

{\em kmo\_dark} & 
Bias, readnoise, dark current, number of bad pixels &
Indicates general status of detector and readout electronics \\

{\em kmo\_flat} &
Lamp efficiency, saturation, signal-to-noise &
Determines whether brightness of lamp is changing, if it is too
bright, and if there were enough exposures \\

 &
Mean \& rms for coefficients of fits to slitlet edges &
Reflects stability of spectra on detector (e.g. 0$^{th}$ \& 1$^{st}$
order coefficients indicate location \& tilt) \\

{\em kmo\_wave\_cal} &
Ar \& Ne lamp efficiencies, saturation, signal-to-noise &
Determines whether brightness of lamps is changing, if they are too
bright, and if there were enough exposures \\

 &
Mean \& rms for coefficients of dispersion solution &
Reflects stability of spectra on detector (e.g. 0$^{th}$ order
coefficients indicate location) \\

 &
Spectral resolution &
Useful for science; a change may indicate intrument is no longer
focussed \\

{\em kmo\_illuminate} &
Inter- and intra-IFU uniformity &
Indicates impact of, and changes in, vignetting \\

{\em kmo\_std\_star} &
Zeropoint, spatial resolution &
Indicates whether instrument throughput has changed \\

\end{tabular}

\section{Data Processing}
\label{sec:dataproc}

Generating fully reduced cubes from the raw KMOS data is an intricate
processes that has to fulfil a number of roles.
For KMOS, the software has been designed using the experience gained
at MPE over more than a decade from both 3D\cite{wei96} and
SINFONI\cite{eis03,bon04,abu06}. It incorporates the perspective of
astronomers as well as programmers.

\subsection{General Requirements}

The detailed processing steps may vary considerably between different
observations, depending on the targets and the science that is being
addressed.
The software must be flexible enough to provide a high quality
reduction for all possible science cases without overwhelming the
astronomer with endless parameters.
For KMOS, this is achieved by making the pipeline modular.
In practice, the pipeline can be run either in a monolithic way (so
that the appropriate modules are automatically executed sequentially)
or controlled on the command line, perhaps using a script.
In the latter case, users can stop at any point, execute their own
code, and then continue.

A high fidelity `science grade' reduction may take some time to
process.
However, the pipeline must also provide rapid
processing for the real-time display.
This is achieved in several ways.
Unncessary steps can be omitted -- for example flatfielding or
telluric correction may not be needed to reconstruct an image of the
target that is good enough for acquisition purposes.
The speed of the reconstruction depends on the algorithm used.
The fastest and simplest, `nearest neighbour', is sufficient for
acquisition.
The most time consuming step is generating the neighbourhood map,
which is the list of pixels on the detector that are closest to each
pixel in the reconstructed cube (see Section~\ref{sec:recon}).
In the context here, re-using the
neighbourhood map can make the reconstruction much faster, providing
the 24 images for the real-time display within a few seconds.

Error propagation, to yield an estimate of the uncertainty in the
final reduced cube, is performed in a fairly simple manner.
An initial estimate of the error is made for each raw frame using the
counts in each pixel, together with the known read noise and gain.
This is then propagated in a standard way (but ignoring cross terms when
adding or multiplying frames).
The same method can also be applied for most of the interpolation
schemes outlined in Section~\ref{sec:recon}, in which the output value
is just a weighted sum of input values.
It is in the combining stage that the situation becomes more complex.
Here, systematics mean it is likely that combining the input noise
estimates will underestimate the true noise.
Instead, if there are enough values, one can derive a better estimate
of the noise directly from their variance.
This requires that each input value has the same weight, which is a
reasonable assumption for near-infrared data taken with the same
exposure times.
Estimating the noise for integrated spectra is also non-trivial, because
of the correlations between pixels.
This has been analysed in some detail and it has been
shown\cite{for09} that the noise does not decrease as one might
naively expect; and that a scaling factor, which depends on the
aperture size, must be applied.

\begin{figure}
\begin{center}
\includegraphics[height=5cm]{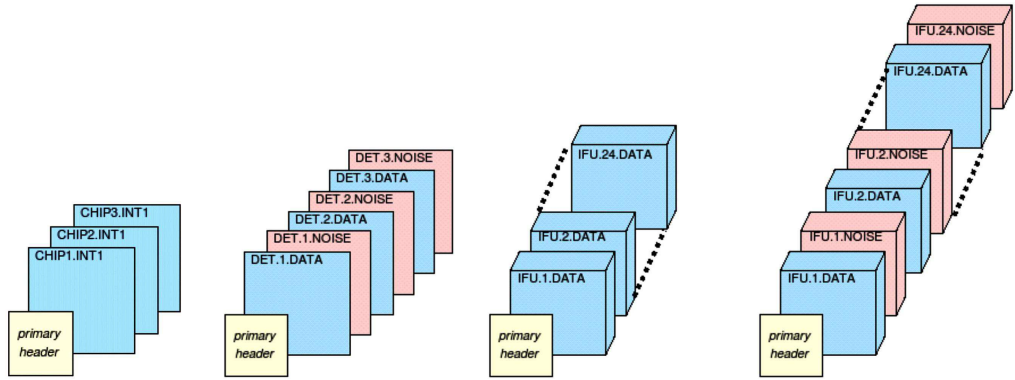}
\end{center}
\caption{ \label{fig:dataformat} 
Overview of the KMOS data format. A keyword EXTNAME in the extension indicates
what data is there. Example string values for this keyword are shown
in the figure.
Far left: initially there are 3
raw frames based on the detector format (i.e. 2k$\times$2k) stored in 3
extensions.
Centre left: once the associated noise is calculated, the additional
frames are interleaved to make 6 extensions.
Centre right: The reconstructed cubes (or collapsed images or extracted
spectra) are each stored in a separate extension.
Far right: interleaving noise cubes yields up to 48 extensions.
}
\end{figure}

The data format itself is an additional complexity for the user,
because of the need to handle, and associate in a single file, either
data from 3 detectors or datacubes for 24 arms.
For this reason, each KMOS file can have multiple extensions, as
illustrated in Fig.~\ref{fig:dataformat}.
For the detector based format (e.g. raw data, dark frames,
flatfields), there will be either three or six extensions depending on
whether noise frames are included;
and each extension will have 2D data.
For data products (e.g. cubes, collapsed images, extracted spectra),
there can be up to 48 extensions; and the data in each extension can
have 1, 2, or 3 dimensions.
For this reason, there is a keyword in each extension that indicates
what data is stored there.
It also  means that it is no longer straightforward for astronomers to
work on the data without modifying their tools to deal with these data
formats.

Simulations are a key aspect of testing the data processing.
In order to make the recipes as robust as possible, we have tested
them using two different sets of simulations.
The first set aims to test the functionality using data that is
representative of KMOS.
These data are generated directly in the 2D detector based format.
As such, it is quick and easy to change anything, for example the
curvature or dispersion.
These simulations are useful for testing different scenarios, and for
developing the recipes since the processing procedures cannot rely on
specific {\em a priori} knowledge about the data format. 
The second set of simulations aim to be much more realistic and
quantitatively resemble KMOS.
Generating them takes time since a 3D source is mapped onto the 2D
detector using the known instrumental properties.
These are important for testing the validity of the reconstruction.
Requiring that the pipeline recipes can process both sets of
simulations leads to robustness and is a good indication that it will
be able to deal with the real KMOS data.

\subsection{Flexure}

Flexure is a potential issue for KMOS since the
instrument itself rotates to maintain the field orientation, and so
the gravity vector is constantly changing.
The impact of flexure is simply to shift the location on the
detector (along both spectral and spatial axes) at which data is
recorded. 
Preliminary estimates, based on models and end-to-end tests, indicate that
within an individual frame, it should not be a problem.
But across a long sequence of frames, in the worst case the shift may
both be as much as about 0.5\,pixel.
At this level, it must be corrected.
Fortunately, along both spectral and spatial axes flexure can be
tracked using the OH emission lines.
As we show below, both spatial and spectral flexure can be fully compensated using
2-pass processing without compromising the data quality.
This is reflected in the workflow given in Fig.~\ref{fig:workflow2}.
 
Uncorrected spectral flexure will result in poor background
subtraction and also broadening of emission lines in the science target.
It can be measured most easily in a reconstructed cube
by comparing the wavelength of several spatially integrated OH lines to a
reference.
A method to do exactly this\cite{dav07} is frequently used on SINFONI data.
With KMOS, we have the advantage that the offset can be fed back into
the calibration data and the cube reconstructed again, thus avoiding
additional interpolations. 

Uncorrected spatial flexure will result in a mismatch between the
flatfield and the data (i.e. some data will be lost), and also a shift
of the target in the field of view.
The former effect can be dealt with by taking flatfields when KMOS is
rotated to several different angles (the number depending on the
severity of the flexure).
The software then needs to select the flatfield taken at the angle
that best matches those of the science observations.
This can therefore be fully compensated, although at the expense of
longer daytime calibration requirements.
The latter effect can be corrected in a similar way to the spectral
flexure, by feeding the measured offset (again most easily derived
from a reconstructed cube) into the calibration data and reconstructing the
cube again.

\subsection{Pipeline Modules and Workflows}

\begin{figure}
\begin{center}
\includegraphics[height=8cm]{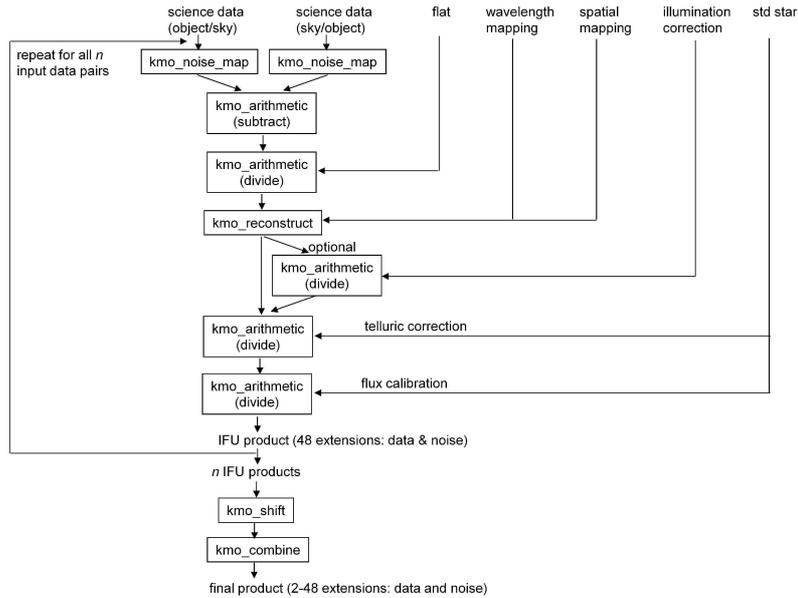}
\end{center}
\caption{ \label{fig:workflow1} 
One possible workflow for the KMOS pipeline. 
This is a standard workflow
in which subtracting the sky is the first step, and in which each cube
is reconstructed before they are shifted and aligned.}
\end{figure} 

The data processing pipeline is constructed from various modules
(recipes) as illustrated in Fig.~\ref{fig:workflow1}.
A user can either execute each recipe in turn, or use the pipeline
recipe to execute them all automatically in sequence.
The function of the main recipes used here are outlined below.
Each of these recipes works on all the extensions in the given data file.\vspace{-3mm}

\begin{description}

\item{\em kmo\_noise\_map} makes an initial estimate of the noise for
  raw frames from the counts, using the known read noise and gain.\vspace{-1mm}

\item{\em kmo\_arithmetic} performs simple mathematical operations on
  cubes using scalars, spectra, images, or other cubes; and on
  images using either scalars or other images. The associated
  noise frames are propagated appropriately. \vspace{-1mm}

\item{\em kmo\_reconstruct} creates a 3D cube from its 2D data using
  the calibration files. The scheme and interpolation methods are
  described in Section~\ref{sec:recon}.\vspace{-1mm}

\item{\em kmo\_shift} performs sub-pixel shifts to align a set of
  cubes on the same WCS. Noise is also propagated.\vspace{-1mm}

\item{\em kmo\_combine} combines cubes that are on the same WCS
  (i.e. have only integer pixel shifts between them). As default,
  only corresponding IFUs are combined; but for the `mapping' mode, all
  IFUs are combined into a single product.\vspace{-1mm}

\end{description}

\begin{figure}
\begin{center}
\includegraphics[height=8cm]{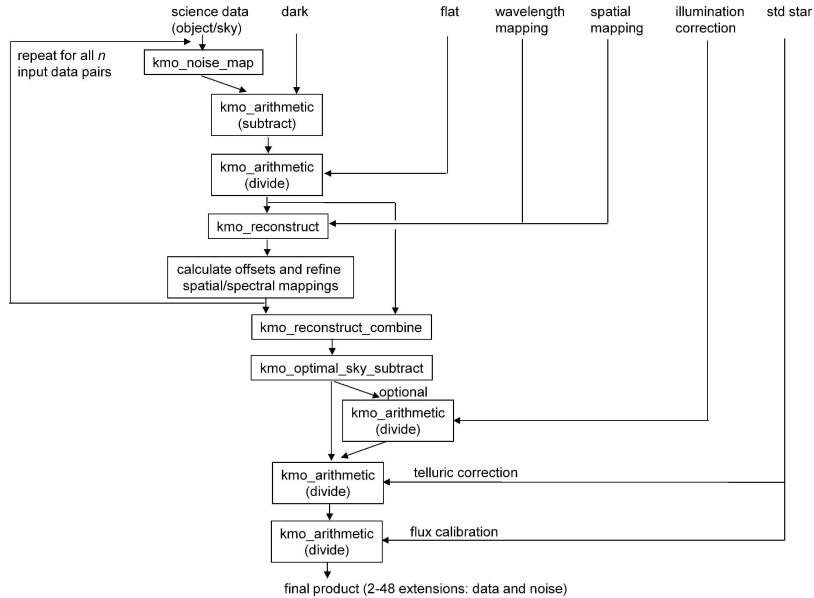}
\end{center}
\caption{ \label{fig:workflow2} 
One possible workflow for the KMOS pipeline. 
This is an advanced workflow in which all the data are combined and
reconstructed simultaneously in a single step, and the sky is
subtracted afterwards 
in an optimal fashion (i.e. minimising OH residuals). In the scheme
shown here, the data are also corrected for spatial and spectral
offsets/flexure, which are derived for each dataset after a
temporary reconstruction step.} 
\end{figure} 

In the second workflow shown in Fig.~\ref{fig:workflow2} there are some
additional recipes which are based on those described above, which
allow one to reconstruct a cube from multiple data sets, or to perform
an optimal sky subtraction.
In addition, there are a number of recipes and tools that are used
only indirectly in the pipeline, but perform tasks that
are frequently required.\vspace{-3mm}

\begin{description}

\item{\em kmo\_stats} calculates a number of useful statistical measures.\vspace{-1mm}

\item{\em kmo\_make\_image} collapses a cube spectrally to create an
  image. One can specify a global wavelength range, and also omit
  specific regions around bright OH lines.\vspace{-1mm}

\item{\em kmo\_fit\_profile} fits various profile forms to a spectral or
  spatial (1D and 2D) profile. This is useful for measuring the
  seeing, the spectral resolution, or aligning compact bright objects.\vspace{-1mm}

\item{\em kmo\_extract\_spec} extracts an integrated spectrum, and can use
  different methods based on apertures or (non-binary) masks.\vspace{-1mm}

\item{\em kmo\_rotate} rotates a cube, which can be useful if KMOS was
  set at a non-zero rotator offset during observations.\vspace{-1mm}

\end{description}

\subsection{Additional Tools}

Handling datacubes can be complex so we provide some extra more
advanced tools that may be useful for
an astronomer but are not used in the pipeline.
These may require significant user interaction in order to optimise
the parameters.\vspace{-3mm}

\begin{description}

\item{\em kmo\_cosmic} is a 3D version of the very effective gradient method
  implemented in LACosmic\cite{dok01} to detect deviant pixels.\vspace{-1mm}

\item{\em kmo\_voronoi} is a spatial binning scheme based on Voronoi
  tessellations that aims to achieve uniform
  signal-to-noise\cite{cap03}. This routine does not affect the
  spectral resolution.\vspace{-1mm}

\item{\em kmo\_extract\_pv} creates a position-velocity diagram from
  the cube, as one would obtain from a 2D longslit spectrum.\vspace{-1mm}

\item{\em kmo\_extract\_moments} derives the kinematics for an
  emission line using both moment calculations, and also by convolving
  the instrumental line profile with a Gaussian (so as to derive the
  intrinsic line properties).\vspace{-1mm}

\item{\em kom\_fit\_continuum} allows the user to fit and subtract the
  continuum from each spectrum in a cube, in order to leave just the
  emission (and absorption) lines.

\end{description}

\section{Data Cube Reconstruction}
\label{sec:recon}

\subsection{Philosophy}

Interpolation is a crucial issue for integral field spectroscopy,
since reconstructing 3D datacubes requires a significant amount of
interpolation\cite{dav08}.
In the classical approach, one needs to correct for bad pixels,
straighten the spectral traces, linearise the dispersion, and finally
align the slitlets (or pixels).
Tuning the wavelength scale to correct for flexures between frames (as
described in Section~\ref{sec:dataproc}) can introduce an additional
interpolation step.
Poor management of the interpolation strategy or poor choice of the
interpolation scheme can degrade the quality of the final data.
For this reason, KMOS makes use of an alternative perspective on the purpose
of calibrations which allows one to view the data reconstruction in a
different way.
In principle this enables one to perform all the interpolation in a
single step while at the same time permitting a far greater
flexibility.
This can be summarised as follows:\vspace{-3mm}

\begin{description}

\item{\em Standard View}:
Calibrations allow one to create the mathematical functions 
(e.g. polynomials) needed to correct the spectral and spatial curvature on
the detector. With this perspective 
(i) bad pixels must be first corrected,  
(ii) the raw data itself sets the pixel scale of the calibrated
product, and
(iii) frames must be individually corrected first before they can be
aligned and combined.\vspace{-1mm}

\item{\em Alternative View}:
Calibrations allow one to create look-up tables which associate each
measured value on the detector with its spectral and spatial location
in the final reconstructed data. This perspective means that 
(i) bad pixels are simply ignored because they do not appear in the look-up
tables, 
(ii) one has the freedom to choose any spectral and spatial sampling
in the processed product, and 
(iii) by including multiple frames in the loop-up table, one can
combine and interpolate all the data simultaneously in one step.\vspace{-1mm}

\end{description}

\begin{figure}
\begin{center}
\includegraphics[height=6cm]{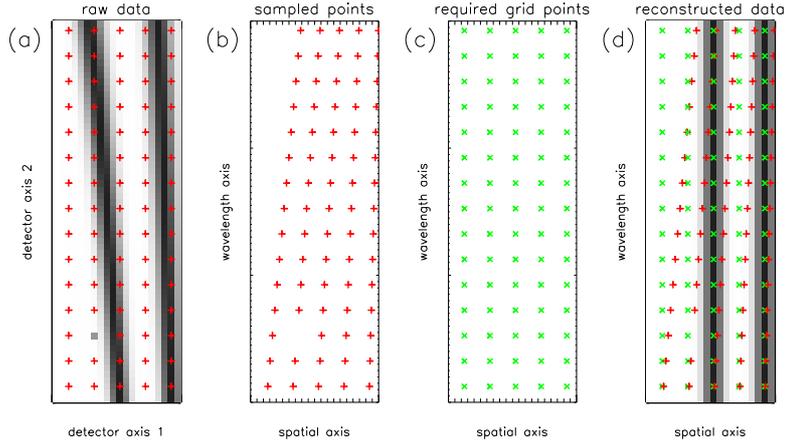}
\end{center}
\caption{ \label{fig:interp} 
Interpolation scheme illustrated in 2-dimensions.
(a) observed data are sampled regularly in the reference frame of the detector.
(b) this sampling is irregular in the reference frame of the
  reconstructed cube; bad pixels can simply be omitted from the set
  of sampled points.
(c) one can freely specify the required gridding
  (i.e. spatial/spectral pixel scale)
  for the reconstructed data; it is independent of the actual sampling.
(d) each required grid point is interpolated from sampled points which
  lie in its local neighbourhood. Any suitable algorithm can be used
  for the interpolation.
}
\end{figure} 

This alternative perspective, which is used for KMOS, is outlined
graphically in Fig.~\ref{fig:interp}.
The most important realisation is that in `detector space' there can
be no concept of a wavelength or spatial axis.
These concepts apply only to the final reconstructed cube.
The detector is nothing more than the medium on which raw data values are
recorded.
The calibrations allow one to assign each measured value on the
detector to a spatial/spectral location in the
reconstructed cube.
Together, these locations provide an irregularly spaced sampling of
that cube.
The aim is thus to reduce the raw data and the calibrations to
a list of values with their associated locations:
\[
\begin{array}{cccc}
value_0, & x_0, & y_0, & \lambda_0 \\
value_1, & x_1, & y_1, & \lambda_1 \\
\vdots & \vdots & \vdots & \vdots \\
value_n, & x_n, & y_n, & \lambda_n \\
\end{array}
\]
Data associated with bad pixels is simply excluded from the list and
so does not contribute to the set of sampled locations.
Creation of this list is the first step.
The second step is to specify the regular sampling -- i.e. the spatial
and spectral pixel size -- that is required for the reconstructed
cube.
The third step is to interpolate each of these regularly gridded
positions from sampled locations in the local neighbourhood.
In a fourth step, one can determine any spectral (or spatial) offsets
in the reconstructed cube and feed these back to create a 
new list with updated locations for each measured value.
One can then re-interpolate the regular grid of points, leading to a
final cube which has been reconstructed in a single interpolation and
which has no offsets.
It is possible to take this scheme further, by adding multiple frames
to the look-up table that lists the data values and their associated
locations. 
In doing so, one needs to include the spatial offsets between the
frames, but this can be determined either from keywords in the frame
headers or by measuring offsets directly from the temporarily
reconstructed cubes of each individual frame.
But ultimately, this allows one to perform a single interpolation that
simultaneously reconstructs and combines (and rotates) multiple frames.
This is the essence of the workflow illustrated Fig.~\ref{fig:workflow2}.

\subsection{Interpolation Methods}

There are several specific intepolation schemes that have been
considered for KMOS. Not all of these will necessarily be implemented.\vspace{-3mm}

\begin{description}
\item{\em Nearest Neighbour} is a fast but approximate method. It may
  be a good choice for noisy data (since it does not degrade the
  signal-to-noise at all) when the highest spatial or spectral
  resolution is not required.\vspace{-1mm}

\item{\em Modified Shepard's Method}\cite{ren88} is a localised
  quadratic inverse distance weighted method. It is a relatively fast
  and accurate method\cite{yan04}. We have implemented a simplified
  version which interpolates from the data itself, rather than from a smooth
  quadratic function that is fit to the data.\vspace{-1mm}

\item{\em Kriging}\cite{cla00} is a distance weighted scheme that is
  optimal in the sense that interpolated values have the smallest
  variance. Its strength is that it takes into account
  correlations between neighbouring pixels based on the seeing and
  spectral resolution, and provides an estimate of the uncertainty.\vspace{-1mm}

\item{\em Drizzling}\cite{fru02} is a linear reconstruction method
  introduced to cope with undersampled 
  data. It is an ideal method to optimise the spatial resolution of
  the combined data when the observations were performed in excellent
  seeing, or the data sets taken at sub-pixel offsets.\vspace{-1mm}

\item{\em Cubic Spline} interpolation is a standard technique. For
  irregularly sampled data (i.e. when reconstructing a cube) it has to
  be applied in 3 successive 1D interpolations. But for shifting or
  rotating data that is already on a regular grid, a single bicubic
  spline interpolation is possible.

\end{description}

\subsection{The Neighbourhood Map}

One step that is implicit in all of the above methods is the creation
of a neighbourhood map.
This is the list of pixels in the detector based format that lie close
to each pixel (voxel) in the reconstructed cube.
For the nearest neighbour, there will be only one detector pixel
for each cube voxel.
But the other methods require lists of 20 or 30 detector pixels for
each voxel in order to perform an appropriate interpolation.
It is the creation of this neighbourhood map that takes most of the
time.
However, the mapping is fixed for any given set calibration frames
(XCAL, YCAL, and LCAL) and so can be re-used.
For acquisition, when speed is of the essence, the cubes will be
reconstructed using a set of master neighbourhood maps that are
updated only occasionally.
By doing so, it will be possible to see the reconstructed data in
quasi-real time.
Similarly, if astronomers are processing their data at home, the
first time the interpolation is performed it will take some time.
However, thereafter the neighbourhood map can be re-used, and
reconstructing additional data sets will be very much faster.

\section{Visualisation}

There are now many tools available for visualising 3D datacubes.
However, the format of the data and the fact that there are 24 cubes
for each exposure mean that most of these tools are not compatible with, or ideal for, KMOS.
QFitsView is a popular and verstile tool, that has been developed at
MPE and has been included in ESO's SciSoft releases for several years.
It is available from
\verb+http://www.mpe.mpg.de/~ott/QFitsView+,
and is now being upgraded to handle KMOS data.
QFitsView already enables simple versions of many detailed analyses to be
performed.
However, it is not yet clear how best these should be applied when
there are multiple datasets within a buffer.
Such issues will be considered on an `as-needs' basis, and the initial
upgrade will address only more essential aspects as follows:\vspace{-3mm}
\begin{itemize}
\item
recognise KMOS files and open all extensions as subsets of a single buffer.\vspace{-1mm}
\item
show a list of the extensions, with their IFU identification as well
as whether the contents refer to are data or noise; allow the user to
select all or some of the datasets and display them.\vspace{-1mm}
\item
simultaneously display a representation of as many spectra, images, or
cubes as are selected, with an appropriate scaling.\vspace{-1mm}
\item
for multiple cubes, enable the user to change the
representation of all datasets simultaneously to be a single
wavelength slice, a combination of slices within a given wavelength
range, or a continuum-subtracted line map.\vspace{-1mm}
\item
save datasets as individual `standard' fitsfiles, i.e. having no
extensions, and compatible with the tools with which most astronomers
are more familiar).
\end{itemize}



\end{document}